# 1/f Noise and the Infrared Catastrophe

Ferdinand Grüneis

Institute for Applied Stochastic
Untersölden 1, 94034 Passau, Germany

Email: Grueneis-IfAS@t-online.de

It is generally assumed that stochastic processes exhibiting 1/f noise are affected with the so-called infrared catastrophe. We present an intermittent stochastic process generating 1/f noise which avoids this problem.

*Keywords:* 1/f noise; noise and fluctuation phenomena.

## 1. Introduction

In physics, situations occur in which an integral diverges. By analogy to the ultraviolet catastrophe taking place for the Rayleigh-Jeans formula of the black-body radiation this is often denoted the infrared catastrophe. The divergence of an integral may have different reason as, for example, divergence in context with Feynman paths because of contributions of objects with very small energy approaching zero.

Another example is the divergence problem of 1/f noise. Many phenomena in physics, biology and even economical systems [1] exhibit fluctuations which can be described by a power spectral density according to $1/f^b$ with an exponent $b$ close to one ($f$ is the frequency). The total power of 1/f noise is proportional to the integral

$$\int_{f_l}^{f_u} \frac{df}{f} = ln\left(\frac{f_u}{f_l}\right)$$

where $f_l$ is a lower and $f_u$ an upper frequency limit of 1/f noise. In case $f_l$ tends to zero, the integral diverges. In resistors or semiconductor materials, for example, 1/f noise is observed down to a lower frequency limit of $10^{-6}$ Hz [2]. The 1/f shape is supposed to go on to even lower frequencies and it is only the observation time (and the stability of recording instruments) which sets a practical limit to $f_l$.

From a mathematical point of view, the divergence problem can be overcome by modifying the power law of *1/f* to *1/f R( f )* where the function *R( f )* is assumed to decrease slowly when $f_l$ tends to zero. Based on this modified power spectrum, Mandelbrot [3] gives a reasonable explanation of the infrared catastrophe. Following Mandelbrot, Chen [4-5] proposes such an augmented function for explaining the propagation of acoustic waves in dissipative media.

In this paper, we investigate an intermittent process which is generating 1/f noise. We show that the spectral power of this process remains finite even if the lower frequency limit $f_l$ approaches zero. We show that there is no need for an augmented function *R( f )* suggested by Mandelbrot; rather, a compensating factor is produced by the intermittent process itself.

## 2. The Intermittent Stochastic Process

An intermittent stochastic process $y(t)$ is characterized by clusters of events separated by distinct intermissions. The intermission is denoted by $\delta$, the cluster duration by $\tau_c$ and the inter event time in a cluster by $\tau_0$. For a schematic plot see Fig. 1. The shape of an event is described by $x(t)$; we regard events with amplitude $A$ and duration $\tau_x$ described by for $0 \leq t < \infty$: $\quad\quad\quad\quad x(t) = A\exp(-t/\tau_x)$.

The Fourier transform is

$$X(f) = A\tau_x/(2\pi i f\tau_x - 1). \tag{1}$$

The events within a cluster are assumed to occur at random. Hence, the probability density function of the inter event time $\tau_0$ is exponentially distributed as[1]

$$p_{\tau_0}(t) = \exp(-t/\tau_0)/\tau_0 \tag{2}$$

applying for $0 \leq t < \infty$.

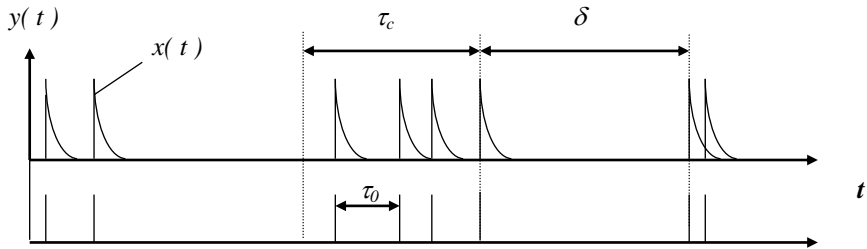

Fig. 1. Schematic plot of an intermittent process $y(t)$. The corresponding point process is shown below.

### 2.1. *The power spectral density of the intermittent stochastic process*

The power spectral density of such an intermittent stochastic process has been derived elsewhere [6]; the one-sided spectrum is given by

$$G_y(f) = \frac{2}{\tau_c + \delta}\left\{ \overline{N}\,\overline{|X(f)|^2} + \overline{|X(f)|}^2 \Phi_{ex}(f) \right\}. \tag{3}$$

$\overline{N}$ is a mean cluster size. Taking (2) into account, the characteristic function of inter event time is obtained by

$$U_{\tau_0} \equiv U_{\tau_0}(f) = 1/(1 - i2\pi f\tau_0)$$

and that of the cluster duration by

$$U_{\tau_c} \equiv U_{\tau_c}(f) = \sum p_m U_{\tau_0}^m(f).$$

Herein $p_m$ is the probability of finding exactly $m$ events in a cluster. Hence, a mean cluster size is

$$\overline{N} = \sum_{m=1}^{N_{max}} p_m m \qquad (4)$$

and the mean duration of a cluster

$$\tau_c = \overline{N}\tau_0 . \qquad (5)$$

---

[1] To avoid an extended mathematical formalism we do not distinguish between a random variable and its mean value.

In terms of the characteristic functions of intermission $U_\delta$, of inter event time and of cluster duration the excess noise function can be expressed by

$$\Phi_{ex}(f) = 2\,Re\left[\frac{(U_\delta - 1)(U_{\tau_c} - 1)}{U_\delta U_{\tau_c} - 1} \frac{U_{\tau_0}}{(1 - U_{\tau_0})^2}\right]. \qquad (6)$$

### 2.2. *Specification of intermission*

The further consideration apply to the case that intermission is statistically equivalent to the cluster duration; this implies that

$$\delta = \tau_c \qquad \text{and} \qquad U_\delta(f) = U_{\tau_c}(f). \qquad (7.\text{a-b})$$

An example for such a choice is given in [7]. In this way, the intermission $\delta$ in (6) is eliminated as an independent statistical variable. Substituting (7.b) into (6), the excess noise function can be transformed to

$$\Phi_{ex}(f) = 2\,Re\left[\frac{U_{\tau_c} - 1}{U_{\tau_c} + 1}\frac{U_{\tau_0}}{(1 - U_{\tau_0})^2}\right].$$

Inserting (7.a) in combination with (5) into (3), the spectrum of the intermittent process can be written as

$$G_y(f) = \frac{1}{\tau_0}\left\{\overline{|X(f)|^2} + \overline{|X(f)|^2}\frac{\Phi_{ex}(f)}{\overline{N}}\right\}. \qquad (8)$$

The first term is shot noise due to the overall occurrence of events; the second term is excess noise originating from fluctuating clusters and intermissions.

### 2.3. *The intermittent process generating $1/f^b$ noise*

An excess noise with a pure 1/f shape has been found numerically for a cluster size distribution

$$p_m \propto m^{-2} \quad \text{for } m = 1, 2, ..., N_{max}. \qquad (9)$$

$N_{max}$ is a maximum cluster size; for the following we assume $N_{max} \gg 1$. The excess noise is excellently approximated by

for $f_l < f < f_u$:
$$\Phi_{ex}(f) \approx \frac{0.3}{f\tau_0}. \qquad (10)$$

The lower and the upper frequency limit $f_l$ and $f_u$ correspond to a maximum and minimum cluster size respectively

$$f_l \approx 1/N_{max}\tau_0 \qquad \text{and} \qquad f_u \approx 1/\tau_0.$$

For a schematic plot see Fig. 2. Hence the scaling range

$$f_u / f_l \approx N_{max} \qquad (11)$$

increases with a maximum cluster size. Substituting (1) and (10) into (8), the spectrum of the intermittent process is expressed by

$$G_y(f) = \frac{1}{\tau_0}\left\{\overline{A^2} + \overline{A}^2 \frac{0.3}{f\,\overline{N}\,\tau_0}\right\}\frac{\tau_x^2}{1+(2\pi f \tau_x)^2}. \qquad (12)$$

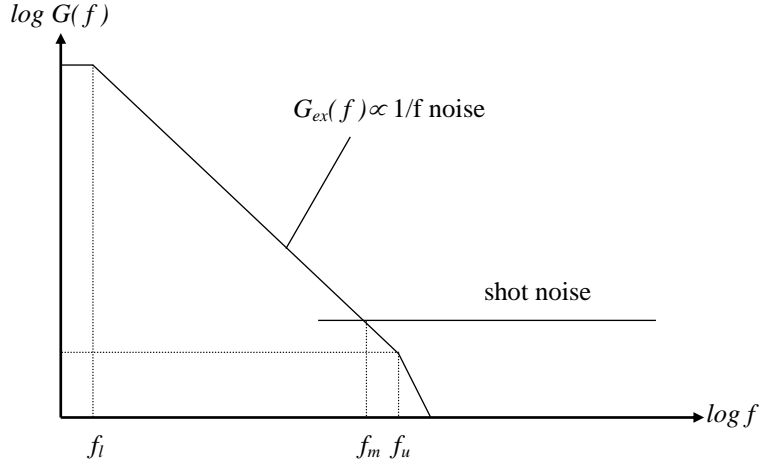

Fig. 2. Schematic double-logarithmic plot of the power spectral density of the intermittent process. $f_l$ is the lower, $f_m$ the intermediate and $f_u$ the upper frequency limit to 1/f noise.

### 2.4. *Power Spectral Balance of the Intermittent Process*

According to (9), a mean cluster size in (4) is approximated by

$$\overline{N} \approx \frac{6}{\pi^2} \sum_{m=1}^{N_{max}} m^{-1} \approx 0.6\, \ln N_{max}.$$

Substituting this into (12), the excess noise can be written as

$$G_{ex}(f) = \frac{\overline{A}^2 \tau_x^2}{2\tau_0^2 \ln N_{max}} \frac{1}{f}. \qquad (13)$$

With the help of (11) it is seen that the excess noise is the lower the larger the scaling range. On the other hand, the shot noise level (= first term in (12)) remains constant. As a consequence, the intermediate frequency at which shot noise is equal to 1/f noise is

$$f_m = \frac{1}{2\tau_0 \ln N_{max}}$$

whereby $\overline{A^2} = \overline{A}^2$ has been assumed. According to (11) $f_m$ is shifted to lower frequencies for an increasing scaling range. The integral over the scaling range of the excess noise in (13) is

$$\int_{f_l}^{f_u} G_{ex}(f) \approx \frac{\overline{A}^2 \tau_x^2}{2\tau_0^2 \ln N_{max}} \ln(f_u/f_l).$$

The compensating factor can be identified as $\ln N_{max}$. Applying (11) we end up with

$$\int_{f_l}^{f_u} G_{ex}(f) \approx \frac{\overline{A}^2 \tau_x^2}{2\tau_0^2}.$$

Hence, the integral over the scaling range of 1/f noise remains finite for an arbitrarily small frequency limit $f_l$.

### 3. Summary and Results

We investigate an intermittent stochastic process which is characterized by clusters of events separated by distinct intermissions. We regard the case that the intermission has the same statistical features like the duration of a cluster; such an intermittent process has found an application for the interpretation of 1/f noise in physical systems [7]. The power spectral density of the intermittent process is regarded for a cluster size distribution which follows a power law as $p_m \propto m^{-2}$. The spectrum contains two contributions: shot noise due to the over all occurrence of events and 1/f noise due to fluctuating clusters and intermissions. 1/f noise scales between a lower frequency limit $f_l$ which correspond to a maximum cluster size and an upper limit $f_u$ corresponding to a minimum cluster size.

Contrary to Mandelbrot [2] and Chen [3-4] we do not introduce an auxiliary function $R(f)$ which should vary very slowly when $f$ tends to zero. Rather a compensating factor is produced by the intermittent process itself. This compensating factor is not dependent on the frequency but on the scaling range of 1/f noise. In this way, the total power over the 1/f noise spectrum remains finite for an arbitrarily large scaling range.